\documentstyle[amssymb,12pt]{article}
\begin{document}
\begin{titlepage}
\thispagestyle{empty}
\begin{flushright} DESY 98-186
\end{flushright}
\vspace{0.5in}
\begin{center}
{\Large \bf Gravitino constraints on models of neutrino masses 
and leptogenesis \\}
\vspace{0.7in}
{\bf  David Delepine$^1$,  Utpal 
Sarkar$^2$\\}
\bigskip
{$^1$ \sl DESY, Notkestrasse 85, D-22607 Hamburg, Germany \\}
\bigskip
{$^2$ \sl Physical Research Laboratory, Ahmedabad 380 009, India \\}
\bigskip
\end{center}
\begin{abstract}

In the supersymmetric extensions of the standard model, neutrino 
masses and leptogenesis requires existence of new particles. We 
point out that if these particles with lepton number violating 
interactions have standard model gauge interactions, then they may 
not be created after reheating because of the gravitino problem. 
This will rule out all existing models of neutrino masses and 
leptogenesis, except the one with right-handed singlet neutrinos.

\end{abstract}
\end{titlepage}

\baselineskip 18pt

Recent announcement of a positive evidence for a neutrino mass from an
observation of the atmospheric neutrino anomaly \cite{atm} and the
supporting evidence from the solar neutrino puzzle \cite{sol} implies an
extension of the standard model. In all the models of neutrino masses
considered so far, require existence of new heavy particles $H$ with a mass $%
M$ and lepton number violating interactions. At low energy this results in
an effective dimension-5 operator ${\cal O}=h^{2}{\frac{1}{M}}\ell _{L}\ell
_{L}\phi \phi $, where $\phi $ is the usual higgs doublet which gives masses
to the quarks and charged leptons. When $\phi $ acquires a vacuum
expectation values ($vev$) to break the electroweak symmetry, the
left-handed neutrinos get small Majorana masses, $m_{\nu }={\frac{(h\langle
\phi \rangle )^{2}}{M}}.$ $H$ could be a $SU(2)_{L}$ singlet or a triplet
and it could be a fermion or a scalar, which gives four possible categories
of models for neutrino masses \cite{dim5}.

In all these models the decays of the new particles $H$ violate lepton
number and can, in general, generate a lepton asymmetry of the universe,
which then gets converted to a baryon asymmetry of the universe before the
electroweak phase transition\cite{ht}. The out-of-equilibrium distribution
of $H$ requires $M$ to be very heavy, particularly if $H$ has got standard
model gauge interactions. In this article we point out that if we constrain
the reheating temperature after inflation by requiring that the universe is
not overpopulated by unacceptably large number of gravitinos, then that
singles out models with right handed neutrinos without any standard model
gauge interactions to be the only consistent class of model for neutrino
masses and leptogenesis. In the following we first discuss the different
models and then discuss the gravitino constraint.

{\sl Right handed neutrinos :} The fermion content of the standard model is
extended to include one right handed neutrinos ($N_\alpha$, $\alpha = 1,2,3$%
) per generation, which are singlets under the standard model gauge group.
The Majorana masses and the Yukawa couplings of the right-handed neutrino
with other leptons are given by, 
\begin{equation}
{\cal L} = M_\alpha \overline{N_\alpha^c} N_\alpha + h_{i \alpha} \overline{%
\ell_{iL}} N_\alpha \phi + h.c. .
\end{equation}
Without loss of generality we assume that the right handed neutrino Majorana
mass matrix is real and diagonal. The lepton number violating scale is then
given by the right-handed neutrino masses, $M_\alpha$. When the higgs
doublet $\phi$ acquires a $vev$, $\langle \phi \rangle = v$, there will be
an induced Dirac mass matrix for the neutrinos, $(m_D)_{i \alpha} = h_{i
\alpha} v$.By the usual see-saw mechanism\cite{seesaw}, the left-handed
neutrinos get a small Majorana mass $M_\nu = - m_D^T {\frac{1 }{M_R}} m_D $.

The decay of the right handed neutrino $N_{\alpha }$ now breaks lepton
number. $CP$ violation comes from an interference of the vertex and
self-energy one-loop diagrams with the tree level diagrams \cite{fy,lg} and
the amount of asymmetry is given by, 
\begin{equation}
\epsilon _{\alpha }\simeq {\frac{3}{16\pi v^{2}}}\sum_{\beta \neq \alpha }{%
\frac{1}{(m_{D}^{\dagger }m_{D})_{\alpha \alpha }}}{\rm Im}\left[
(m_{D}^{\dagger }m_{D})_{\alpha \beta }^{2}\right] {\frac{M_{\alpha }}{%
M_{\beta }}}
\end{equation}
when $M_{\alpha }<M_{\beta }$ and $N_{\alpha }$ is the lightest right handed
neutrino. The decay of $N_{\alpha }$ can generate this amount of lepton
asymmetry of the universe, if it satisfies the out-of-equilibrium condition, 
$\Gamma _{N_{\alpha }}={\frac{h_{i\alpha }^{2}}{8\pi }}M_{\alpha }<H= 1.7%
\sqrt{g_{*}}{\frac{T^{2}}{M_{Pl}}}~~~~~{\rm at}~~T=M_{\alpha }$, where, $H $
is the Hubble constant, $g_{*}$ is the effective number of helicity states, $%
M_{Pl}$ is the Planck scale.

{\sl Triplet higgs :} It is possible to extend the standard model to include 
$SU(2)_L$ triplet higgs scalar fields ($\xi_a \equiv (1,3,-1)$, $a=1,2$ two
of them are required for $CP$ violation), whose relevant interactions are, 
\begin{equation}
{\cal L} = M_a \xi_a^\dagger \xi_a + f_{ij} \xi_a^\dagger \ell_i \ell_j +
\mu \xi_a \phi \phi + h.c.
\end{equation}
We choose $M_a$ to be real and diagonal. The parameter $\mu$ has a dimension
of mass and is of the order of $M_a$, which is the only scale in this model
other than the electroweak symmetry breaking scale. Lepton number is
explicitly broken at this scale, but the triplet higgs acquires a very tiny $%
vev$, which gives a Majorana mass to the neutrinos \cite{ma,triplet}, 
$M_\nu = f_{ij} <\xi_a>
= - f_{ij} {\frac{ \mu <\phi>^2 }{M_a^2 }}$ . Lepton number violation comes
from the decays of the triplet higgs, $\xi_a$. In this case there are no
vertex corrections to these decay processes. The one loop self-energy
diagrams interfere with the tree level decays to give $CP$ violation. If $%
M_2 > M_1$ and $\xi_1$ decays away from thermal equilibrium, {\it i.e.}, $%
\Gamma_1 = {\frac{1 }{8 \pi M_1}} ( \mu_1 \mu_1^* + M_1 M_1 \sum_{k,l}
f^*_{1kl} f_{1kl} ) < H = 1.7 \sqrt{g_*} {\frac{T^2 }{M_{Pl}}} ~~~~{\rm at}%
~~T = M_1$, then a lepton asymmetry will be generated \cite{ma}, 
\begin{equation}
{\frac{n_L }{s}} \simeq {\frac{{Im \left[ \mu_1 \mu_2^* \sum_{k,l} f_{1kl}
f_{2kl}^* \right]} }{{24 \pi^2 g_* (M_1^2 - M_2^2)}}} \left[ {\frac{{\ M_1} 
}{\Gamma_1}} \right].
\end{equation}

{\sl Triplet Majorana fermions :} One can also extend the standard model to
include a $SU(2)_L$ triplet fermions, whose large Majorana masses breaks
lepton number. For all practical purposes they behave similar to those like
the right handed neutrinos and give neutrino masses through see-saw
mechanism \cite{tripletfermion}. Their decay can generate a lepton and hence
baryon asymmetry of the universe in the same way as that of a singlet right
handed neutrino.

{\sl Radiative models :} It is possible to write down the effective
dimension-5 operator with $SU(2)_L$ singlet field ($\chi_a \equiv (1,1,-1)$, 
$a=1,2$, two of these fields are required to get $CP$ violation) if there
are at least two higgs doublets ($\phi_a, a=1,2$). The neutrino masses
originate from radiative diagrams and hence are naturally small \cite{zee}.
The charged singlet scalar ($\chi$) have couplings to two higgses and also
to two leptons of two different generations, thereby breaking the lepton
number explicitly, 
\begin{equation}
{\cal L} = M^\chi_a \chi_a^\dagger \chi_a + f^\chi_{aij} \chi_a^\dagger
\ell_i \ell_j + \mu_\chi \chi_a \phi_1 \phi_2 + h.c.
\end{equation}
Unlike the triplet scalar case, $\chi$ cannot acquire a $vev$ and hence
neutrino masses can come only radiatively. Although this model has been
studied in details for neutrino masses and has been found to have several
interesting features in the context of the present experiments \cite{zee1},
the question of leptogenesis is only recently being investigated \cite{prep}%
. Without going into the details we make here a few general comments about
this model.

Since the decays of this charged scalar violates lepton number in a similar
way to that of the triplet scalars, a lepton asymmetry can be generated in
the same way. The amount of asymmetry generated is also similar, except that
the parameters involved are now changed. The main point we like to stress
here is that the couplings of $\chi _{a}$ violate lepton number at a very
high scale. Since these singlets are charged, they also suffer from the same
problem as the triplets that it interacts very fast with the standard model
gauge bosons. So all the constraints coming due to the gauge interactions of
the particles whose decay generates an asymmetry will also be applicable to
this case.

Among all the above four classes of models for neutrino masses, only models
with a singlet right-handed neutrino does not have any standard model gauge
interaction. In all the three other classes of models, the new particles
whose interactions break lepton number, transform non-trivially under the
standard model. We shall next study the consequences of the standard model
gauge interaction on the generation of a lepton asymmetry of the universe
when these particles decay.

For simplicity, we shall consider a couple of generic heavy scalar $%
H_{a},a=1,2$, which couples to the standard model gauge bosons through gauge
interactions. In a supersymmetric model, the corresponding superpartner will
have similar gauge interactions with the gauginos and hence will suffer from
the same problem. The generic scalar $H_{a}$ could be $\xi _{a}$ as in the
triplet-higgs model, in which case it will interact with the $SU(2)_{L}$ and 
$U(1)_{Y}$ gauge bosons, or it could be $\chi _{a}$ as in the Zee-type
radiative models, in which case it will interact only with the $U(1)_{Y}$
gauge bosons. For the generation of a lepton asymmetry of the universe we
assume that the relevant part of the lagrangian is similar to that of
eqn(6). We choose $M_{a}^{h}$ to be real and diagonal and $\mu ^{h}\lesssim
M_{a}^{h}$. Decay of $H_{a}$ into two leptons and two higgs together violate
lepton number. The tree level and the one loop self energy diagrams
interfere to generate a $CP$ asymmetry $\eta $. We shall also assume that $%
M_{1}^{h}<M_{2}^{h}$, so that first $M_{2}^{h}$ decays and then the decay of 
$M_{1}^{h}$ generates the lepton asymmetry of the universe.

The evolution of lepton number ($n_L = n_\ell - n_{\ell^c}$) is given by the
Boltzmann equation \cite{fry}, 
\begin{equation}
{\cal D} n_L = \eta \Gamma_H [n_H - n_H^{eq} ] - \left( \frac{n_L}{n_\gamma}
\right) n_H^{eq} \Gamma_H -2 n_\gamma n_L \langle \sigma_L |v| \rangle ,
\label{nl}
\end{equation}
where, the operator ${\cal D} \equiv \left[ \frac{{\rm d}}{{\rm d}t} + 3 H
\right]$; $n_H^{eq}$ is the equilibrium distribution of $H_1$ given by $%
n_{H}^{eq}=\frac{TM_{1}^{h2}}{2\pi^{2}}K_{2}(\frac{M_{1}^{h}}{T})$; $\Gamma_H
$ is the thermally-averaged decay rate of $H_a$; $n_\gamma$ is the photon
density and $\langle\sigma_L |v|\rangle$ is the thermally-averaged lepton
number violating scattering cross section.

The number density ($n_H$) of $H_1$ satisfies the Boltzmann equation, 
\begin{equation}
{\cal D} n_H = - \Gamma_H (n_H - n_H^{eq}) + ({n_H}^2 - {n_H^{eq}}^2)
\langle \sigma_H |v| \rangle .  \label{nN}
\end{equation}
The second term on the right is the lepton number {\sl conserving}
thermally-averaged $H_1^\dagger + H_1 \to W_L + W_L$ scattering cross
section of the heavy particles $H_1$. It is instrumental in initially
equilibrating the number density of $H_1$ (which will also erase any lepton
asymmetry created during the decay of $H_2$) but it also prevents departure
from equilibrium for $H_1$, which depletes the generated lepton asymmetry if
this interaction becomes comparable to the expansion rate of the universe.

To solve these equations, we use the dimensionless variable $x={M_{1}^{h}}/{T%
}$, and normalize the particle density by the entropy density, $Y_{i}={n_{i}}%
/{s}$, so $t={x^{2}}/{2H(x=1)}$. We also define the parameters $K\equiv
\gamma \Gamma _{H(x=1)}/H_{(x=1)}$; $K_{H}\equiv \gamma _{H}\langle \sigma
_{H}|v|\rangle _{(x=1)}/H_{(x=1)}$; $\gamma =\Gamma _{H}/\Gamma _{H(x=1)}$; $%
\gamma _{H}=\langle \sigma _{H}|v|\rangle /\langle \sigma _{H}|v|\rangle
_{(x=1)}$ and $\gamma _{s}=n_{\gamma }\langle \sigma _{L}|v|\rangle
/H_{(x=1)}$. If the out-of-equilibrium condition is satisfied ($K\ll 1$),
the final lepton asymmetry is given by $n_{L}\sim \eta /g_{*}$, if there is
no suppression due to the scattering. Since we are interested in the maximum
amount of possible lepton asymmetry in this scenario, we shall always
consider the couplings of $H_{a}$, which satisfies this out-of-equilibrium
condition, $K\ll 1$ at $T\sim M_{1}^{h}$. However, we do not have this
freedom for the scattering processes, since the standard model gauge
coupling constants are involved. We solve these two nonlinear equations
numerically. To get an unsuppressed lepton asymmetry we work with $K\ll 1$.
If we ignore the lepton number conserving gauge interactions of $H_{1}$, the
lepton asymmetry is now given by $n_{L}\sim \eta /g_{*}$. At temperatures of
interest, {\it i.e.}, after the inflation, the restrictions imposed on the
couplings of $H_{a}$ by $K\ll 1$ will not allow $\eta $ to be larger than $%
\sim O(10^{-5})$.

For our analysis we shall thus assume, $K\ll 1$ at $T\sim M_{1}^{h}$ and $%
\eta <10^{-5}$. Taking the $SU(2)_{L}$ gauge coupling constant to be given
by the GUT coupling constant at the GUT scale, we include the effect of the
lepton number conserving gauge interactions of $H_{1}$ and present it in
figure 1.
\input{epsf.sty} 
\begin{figure}[t]
\leavevmode
\par
\begin{center}
\mbox{\epsfxsize=15.cm\epsfysize=10.cm\epsffile{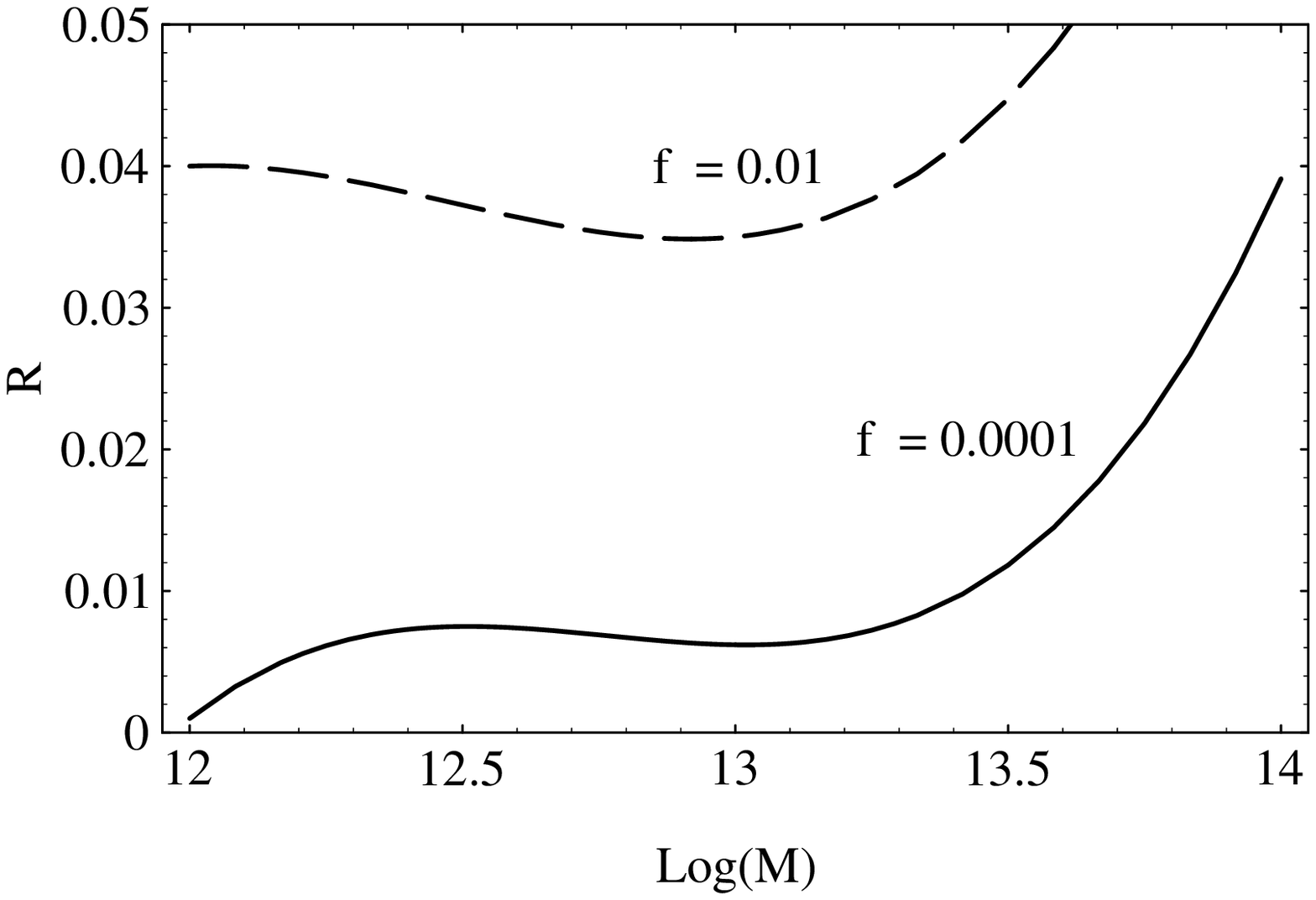}}
\end{center}
\caption{Lepton asymmetry of the universe for different masses of $H_{1}$,
when effects of gauge interaction is included. R is defined as $R= \frac{%
(n_{L}/s)_{with~~H_{1}^{\dagger }+H_{1}\to W_{L}+W_{L}}}{%
(n_{L}/s)_{without~~H_{1}^{\dagger }+H_{1}\to W_{L}+W_{L}}}$ and $%
f=f_{ij}^{h}$}
\label{fig1}
\end{figure}

For very high $M_{1}^{h}$ masses around the GUT scale, when $K_{H}\ll K$ the
lepton asymmetry is given by $n_{L}\sim \eta /g_{*}$. As we lower $M_{1}^{h}$
masses, the scattering processes become comparable to the expansion rate of
the universe and start depleting the amount of lepton asymmetry. For $%
K_{H}\sim K$, we already get a suppression by two orders of magnitude. So,
for the allowed value of $\eta \leq 10^{-5}$ the lowest possible $H_{1}^{h}$
mass for the generation of enough lepton asymmetry of the universe becomes 
\begin{equation}
M_{1}^{h}>O(10^{12})\mathrel{\rm GeV}.  \label{t12}
\end{equation}

The above conclusion will also be true for the triplet Majorana fermion
models, where the lepton number violation in induced by the Majorana mass of
the Majorana triplet fermion $T$. The gauge interactions will then induce a
lepton number conserving scattering process $T^{\dagger }+T\to
W_{L}\rightarrow T^{\dagger }+T$, which will deplete the lepton asymmetry.

We have to keep in mind that leptogenesis can occur only after the end of
inflation. In supersymmetric theories, the thermal production of massive
gravitinos restricts the beginning of the radiation-dominated era following
inflation except when the gravitino is very light\cite{pagels}. After the
inflation a large number of gravitinos are produced, which interact very
weakly. The late decays of unstable gravitinos can then modify the
abundances of light elements causing inconsistency with observation. In the
other hand, stable gravitinos may overclose the universe. This imposes a
upper bound on the reheating temperature $T_{RH}$~~\cite{linde,grav,ab,moroi}%
.

In the case of stable gravitinos, a limit on the $T_{RH}$ can be derived
from the closure limit of the universe\cite{plasma}, 
\begin{equation}
T_{RH}\leq 10^{10}\mathrel{\rm GeV} \times (\frac{m_{3/2}}{100 \mathrel{\rm
GeV}}) \times (\frac{1 \mathrel{\rm TeV}}{m_{\tilde{g}}(\mu)})^{2}
\label{stable}
\end{equation}
with $m_{3/2}$ is the mass of the gravitino and $m_{\tilde{g}}(\mu)$ is the
running mass of the gluino. An upper bound on the gravitino mass also arises
when we are taking into account the effect of the decays of the
next-to-lightest superparticle (NSP) on primordial nucleosynthesis. For
instance, it was shown in Ref.\cite{plasma} than for a $m_{3/2}=1...10^{3}$
GeV and $80$ GeV $<$ $m_{NSP}<300$ GeV, a $T_{RH}$ as large as $10^{8-11}$
GeV could be reached.

In the case of unstable gravitinos, the upper bound on the $T_{RH}$ depends
on the $m_{3/2}$. Essentially, one gets the followings constraints from
primordial nucleosynthesis\cite{moroi2}

\begin{equation}
T_{RH}\leq 10^{9}\mathrel{\rm GeV}~~~~~~~~~~~~~~~~~~~~~m_{3/2}< 1%
\mathrel{\rm TeV}  \label{unstable1}
\end{equation}
\begin{equation}
T_{RH}\leq 10^{12}\mathrel{\rm GeV}~~~~~~~1\mathrel{\rm TeV} < m_{3/2}< 5 %
\mathrel{\rm TeV}  \label{unstable2}
\end{equation}

It has recently been shown that after the reheating temperature ($T_{RH}$),
inflaton decay can produce particles as heavy as $\sim 10^{3}T_{RH}$ with
sufficiently large abundances \cite{riotto}. In that case, they get a
suppression factor coming from the annihilation cross section of the heavy
particles, which is the scattering term in the Boltzmann equation (\ref{nN}%
). Thus the suppression factor is similar to the suppression due to the
departure of the scattering process from equilibrium. In other words,
particles with mass $10$ times $T_{RH}$ will have an abundance about $S\sim
O(10^{-3})$ times less than the equilibrium abundance, which can generate a
lepton asymmetry which is less by a factor of $S$. So, even for the stable
gravitino, when the bound on the reheating temperature is given by eqn(\ref
{stable}), it will not be possible to generate enough lepton asymmetry in
these scenarios, where the lepton number violating particle have got
standard model gauge interactions.

In the left-right symmetric models \cite{lr}, both the right-handed
neutrinos and the triplet higgs are present. However, in most of the
realistic models the triplet higgs are much heavier than the right handed
neutrinos and a lepton asymmetry is generated when the lightest right handed
neutrino decay. In SO(10) GUT the left-right symmetric models are naturally
embedded, in which it has been shown that the mass of the lightest right
handed neutrino comes out to be just consistent with the gravitino bound and
on the other hand the amount of generated lepton asymmetry is also as
required \cite{buch}. This may indicate that the recent positive evidence of
the neutrino masses may actually directing us towards a supersymmetric $%
SO(10)$ GUT.

In summary, we point out that all the supersymmetric models of neutrino
masses, except for the one with singlet right handed neutrinos and
left-right symmetric models, may not be able to generate enough lepton
asymmetry of the universe consistently with the gravitino bound on the
reheating temperature in inflationary universe. So see-saw mechanism of the
neutrino masses with right handed singlet neutrinos and left-right symmetric
models, such as supersymmetric $SO(10)$ or superstring inspired $E(6)$ GUTs,
then becomes the most preferred solution for neutrino masses and
leptogenesis.

\vskip .2in

{\bf Acknowledgements} We thank Prof. W. Buchm\"{u}ller for useful comments
and discussions.


\begin{thebibliography}{99}
\bibitem{atm}  Super-Kamiokande Collaboration : Y. Fukuda {\em et al}, Phys.
Rev. Lett. {\bf 81}, 1562 (1998); hep-ex/9805006; Phys. Lett. {\bf B433}, 9
(1998); T. Kajita, hep-ex/9810001.

\bibitem{sol}  Super-Kamiokande Collaboration: Y. Fukuda {\em et al}, Phys.
Rev. Lett. {\bf 81}, 1158 (1998); Talk by Y. Suzuki at Neutrino'98,
Takayama, Japan (1998).

\bibitem{dim5}  E. Ma, hep-ph/9807386, Phys. Rev. Lett. {\bf 81}, 1171
(1998) (hep-ph/9807386).

\bibitem{ht}  S. Yu. Khlebnikov and M.E. Shaposhnikov, Nucl. Phys. {\bf B 308%
}, 885 (1988); J.A. Harvey and M.S. Turner, Phys. Rev. {\bf D 42}, 3344
(1990).

\bibitem{seesaw}  M. Gell-Mann, P. Ramond and R. Slansky, in {\it %
Supergravity}, eds. P. van Nieuwenhuizen and D. Freedman, (North-Holland,
1979) p.315; T. Yanagida, in {\it Proc. Workshop on Unified Theories and
Baryon Number in the Universe}, eds. A. Sawada and A. Sugamoto (KEK, 1979),
p.95; R.N. Mohapatra and G. Senjanovi\'{c}, Phys. Rev. Lett. {\bf 44}, 912
(1980).

\bibitem{fy}  M. Fukugita and T. Yanagida, Phys. Lett. {\bf B 174}, 45
(1986).

\bibitem{lg}  P. Langacker, R. Peccei and T. Yanagida, Mod. Phys. Lett. {\bf %
A 1}, 541 (1986); A. Acker, H. Kikuchi, E. Ma and U. Sarkar, Phys. Rev. {\bf %
D 48}, 5006 (1993); M. Flanz, E.A. Paschos and U. Sarkar, Phys. Lett. {\bf B
345}, 248 (1995); L. Covi, E. Roulet and F. Vissani, Phys. Lett. {\bf B 384}%
, 169 (1996); M. Flanz, E.A. Paschos, U. Sarkar and J. Weiss, Phys. Lett. 
{\bf B 389}, 693 (1996); W. Buchmuller and M. Pl\"{u}macher, Phys. Lett. 
{\bf B 389}, 73 (1996); Phys. Lett. {\bf 431} (1998) 354; A. Pilaftsis,
Phys. Rev. {\bf D 56}, 5431 (1997).

\bibitem{ma}  E. Ma and U. Sarkar, Phys. Rev. Lett. {\bf 80}, 5716 (1998).

\bibitem{triplet} C. Wetterich, Nucl. Phys. {\bf B 187}, 343 (1981);
 G.Lazarides, Q. Shafi and C. Wetterich, Nucl. Phys. {\bf B181}, 287
 (1981); R.N. Mohapatra and G. Senjanovic, Phys. Rev. {\bf D 23}, 165 
 (1981); R. Holman, G. Lazarides and Q. Shafi, Phys. Rev. {\bf D27},
 995 (1983); G. Lazarides and Q. Shafi, report no hep-ph/9803397. 

\bibitem{tripletfermion}  For instance, R. Foot, H. Lew, X.-G. He, and G.C.
Joshi, Z. Phys.{\bf C 44}, 441 (1989)

\bibitem{zee}  A. Zee, {\ Phys. Lett. }{\bf B 93}, 389 (1980).

\bibitem{zee1}  A. Smirnov,talk given at the 28th International Conference
on High Energy Physics (1996),Warsaw,Poland (hep-ph/9611465); N. Gaur,
A.Ghosal, E. Ma, P.Roy, Phys. Rev {\bf D 58}, 7301 (1998).

\bibitem{prep}  E. Ma, M. Raidal and U. Sarkar, in preparation.

\bibitem{fry}  J.N. Fry, K.A. Olive and M.S. Turner, Phys. Rev. Lett. {\bf 45%
}, 2074 (1980); Phys. Rev. {\bf D 22}, 2953 (1980); Phys. Rev. {\bf D 22},
2977 (1980); E.W. Kolb and S. Wolfram, Nucl. Phys. {\bf B 172}, 224 (1980).

\bibitem{pagels}  H. Pagels, J.R. Primack, Phys. Rev. Lett. {\bf 48}, 223
(1982).

\bibitem{linde}  M.Yu. Khlopov, A. D. Linde, Phys. Lett. {\bf B 138}, 265
(1984).

\bibitem{grav}  J. Ellis, J. Kim and D.V. Nanopoulos, Phys. Lett. {\bf B145}%
, 181 (1984).

\bibitem{ab}  J. Ellis, G.B. Gelmini, J.L. Lopez, D.V. Nanopoulos and S.
Sarkar, Nucl. Phys. {\bf B 373}, 399 (1992).

\bibitem{plasma}  M. Bolz, W. Buchm\"{u}ller and M. Plumacher,
hep-ph/9809381.

\bibitem{moroi}  M. Kawasaki, T. Moroi, Progr. Theor. Phys. {\bf 93}
(1995)879; T. Moroi, Ph.D. Thesis (hep-ph/9503210)

\bibitem{moroi2}  E. Holtmann, M. Kawasaki, K. Kohri and T. Moroi,
hep-ph/9805402

\bibitem{riotto}  D.J.H. Chung, E.W. Kolb and A. Riotto, hep-ph/9809453.

\bibitem{lr}  J.C. Pati and A. Salam, Phys. Rev. {\bf D 10}, 275 (1974);
R.N. Mohapatra and J.C. Pati, Phys. Rev. {\bf D 11}, 566 (1975); R.N.
Mohapatra and G. Senjanovic, Phys. Rev. {\bf D 12}, 1502 (1975); R.E.
Marshak and R.N. Mohapatra, Phys. Rev. Lett. {\bf 44}, 1316 (1980).

\bibitem{buch}  M.Pl\"{u}macher, Nucl. Phys. {\bf B 530}, 207 (1998).
\end{thebibliography}
\end{document}